\input phyzzx
 
 
\input tables

\def\vfootnote#1{\Vfootnote{$^{#1}$}}
\def\footnote#1{\attach{#1}\vfootnote{#1}}

\def\foot{\attach\footsymbolgen\vfootnote{\footsymbol}}
\let\footsymbol=\star
 
\def\Pade{Pad$\acute{\rm e}$\chkspace}
%
%
%
%
 
\catcode`@=11
\def\chkspace{%
  \relax   
  \begingroup\ifhmode\aftergroup\dochksp@ce\fi\endgroup}
\def\dochksp@ce{%
  \unskip              
  \futurelet\chkspct@k\d@chkspc  
}
\def\d@chkspc{%
  \let\nxtsp@ce=\relax
  \ifx\chkspct@k.\else     
    \ifx\chkspct@k,\else
      \ifx\chkspct@k;\else
        \ifx\chkspct@k!\else
          \ifx\chkspct@k?\else
            \ifx\chkspct@k:\else
              \ifx\chkspct@k)\else
              \ifx\chkspct@k(\else
                \ifx\chkspct@k]\else
                  \ifx\chkspct@k-\else
                    \ifx\chkspct@k\egroup\else  
                      \let\nxtsp@ce=\put@space  
                    \fi
                  \fi
                \fi
              \fi
              \fi
            \fi
          \fi
        \fi
      \fi
    \fi
  \fi
  \nxtsp@ce
}
\def\put@space{$\;$}
\catcode`@=12
 
\def\ra{{$\rightarrow$}\chkspace}
\def\etal{{\it et al.}\chkspace}

\def\adhoc{{\it ad hoc}\chkspace}

\def\eg{{\it eg.}\chkspace}

\def\apriori{{\it a priori}\chkspace}
\def\apost{{\it a posteriori}\chkspace}
 
\def\ep{{e$^+$e$^-$}\chkspace}
\def\epa{{e$^+$e$^-$ annihilation}\chkspace}

\def\qu{\quad}

\def\gluino{\relax\ifmmode \tilde{g} \else $\tilde{g}$ \fi\chkspace}

\def\bb{\relax\ifmmode {\rm b}\bar{\rm b}
       \else ${\rm b}\bar{\rm b}$ \fi\chkspace}
\def\cc{\relax\ifmmode {\rm c}\bar{\rm c}
       \else ${\rm c}\bar{\rm c}$ \fi\chkspace}
\def\tt{\relax\ifmmode {\rm t}\bar{\rm t}
       \else ${\rm t}\bar{\rm t}$ \fi\chkspace}

\def\qqg{\relax\ifmmode {\rm q}\overline{\rm q}{\rm g}
\else q$\overline{\rm q}$g \fi\chkspace}

\def\afb{\relax\ifmmode A_{FB} \else
{{$A_{FB}$}}\fi\chkspace}
\def\afbb{\relax\ifmmode A_{FB}^b \else
{{$A_{FB}^b$}}\fi\chkspace}
\def\pafb{\relax\ifmmode \tilde{A}_{FB} \else
{{$\tilde{A}_{FB}$}}\fi\chkspace}
\def\pafbb{\relax\ifmmode \tilde{A}_{FB}^b \else
{{$\tilde{A}_{FB}^b$}}\fi\chkspace}
 
\def\pafbzo{\relax\ifmmode \tilde{A}_{FB}|_{O(0)} \else
{{$\tilde{A}_{FB}|_{O(0)}$}}\fi\chkspace}
\def\pafbfo{\relax\ifmmode \tilde{A}_{FB}|_{\oalp} \else
{{$\tilde{A}_{FB}|_{\oalp}$}}\fi\chkspace}
\def\pafbso{\relax\ifmmode \tilde{A}_{FB}|_{\oalpsq} \else
{{$\tilde{A}_{FB}|_{\oalpsq}$}}\fi\chkspace}
\def\pafbto{\relax\ifmmode \tilde{A}_{FB}|_{\oalpc} \else
{{$\tilde{A}_{FB}|_{\oalpc}$}}\fi\chkspace}
 
\def\pafbbzo{\relax\ifmmode \tilde{A}_{FB}^b|_{O(0)} \else
{{$\tilde{A}_{FB}^b|_{O(0)}$}}\fi\chkspace}
\def\pafbbfo{\relax\ifmmode \tilde{A}_{FB}^b|_{\oalp} \else
{{$\tilde{A}_{FB}^b|_{\oalp}$}}\fi\chkspace}
\def\pafbbso{\relax\ifmmode \tilde{A}_{FB}^b|_{\oalpsq} \else
{{$\tilde{A}_{FB}^b|_{\oalpsq}$}}\fi\chkspace}
\def\pafbbto{\relax\ifmmode \tilde{A}_{FB}^b|_{\oalpc} \else
{{$\tilde{A}_{FB}^b|_{\oalpc}$}}\fi\chkspace}
 
\def\afbo0{\tilde{A}_{FB}|_{O(0)}}
\def\afbo1{\tilde{A}_{FB}|_{\oalp}}
\def\afbo2{\tilde{A}_{FB}|_{\oalpsq}}
\def\afbo3{\tilde{A}_{FB}|_{\oalpc}}
 
\def\lam{\relax\ifmmode \Lambda_{\overline{MS}}
       \else {{$\Lambda_{\overline{MS}}$}}\fi\chkspace}
\def\lamuds{\relax\ifmmode \Lambda^{(3)}_{\overline{MS}}
       \else {{$\Lambda^{(3)}_{\overline{MS}}$}}\fi\chkspace}
\def\lamudsc{\relax\ifmmode \Lambda^{(4)}_{\overline{MS}}
       \else $\Lambda^{(4)}_{\overline{MS}}$\fi\chkspace}
\def\lamudscb{\relax\ifmmode \Lambda^{(5)}_{\overline{MS}}
       \else $\Lambda^{(5)}_{\overline{MS}}$\fi\chkspace}

\def\alp{\relax\ifmmode \alpha_s\else $\alpha_s$\fi\chkspace}
\def\alpbar{\relax\ifmmode \overline{\alpha_s}
       \else $\overline{\alpha_s}$\fi\chkspace}
\def\alpmz{\relax\ifmmode \alpha_s(M_Z)\else $\alpha_s(M_Z)$\fi\chkspace}
\def\alpmzsq{\relax\ifmmode \alpha_s(M_Z^2)
       \else $\alpha_s(M_Z^2)$\fi\chkspace}
 
\def\oalp{\relax\ifmmode O(\alpha_s)\else{{O($\alpha_s$)}}\fi\chkspace}
\def\oalpsq{\relax\ifmmode O(\alpha_s^2)
           \else{{O($\alpha_s^2$)}}\fi\chkspace}
\def\oalpc{\relax\ifmmode O(\alpha_s^3)
           \else{{O($\alpha_s^3$)}}\fi\chkspace}
\def\oalpf{\relax\ifmmode O(\alpha_s^4)
           \else{{O($\alpha_s^4$)}}\fi\chkspace}

\def\plb{Phys. Lett.\chkspace}

\def\prl{Phys. Rev. Lett.\chkspace}
\def\prd{Phys. Rev.\chkspace}
\def\zpc{Z. Phys.\chkspace}

\def\z0{{$Z^0$}\chkspace}
\def\Dst{\relax\ifmmode {\rm D}^* \else {D$^*$}\fi\chkspace}
\def\Dpl{\relax\ifmmode {\rm D}^+ \else {D$^+$}\fi\chkspace}
\def\D0{\relax\ifmmode {\rm D}^0 \else {D$^0$}\fi\chkspace}
\def\Kst{\relax\ifmmode {\rm K}^* \else {K$^*$}\fi\chkspace}
\def\K0{\relax\ifmmode {\rm K}^0_s \else {K$^0_s$}\fi\chkspace}
\def\Kpl{\relax\ifmmode {\rm K}^+ \else {K$^+$}\fi\chkspace}
\def\Kstz{\relax\ifmmode {\rm K}^{*0} \else {K$^{*0}$}\fi\chkspace}
 
\overfullrule=0pt
 \hsize=6in
   \vsize=8.5in
 \voffset 0.5truecm
\baselineskip = 15pt
 
\overfullrule=0pt
 \linepenalty=100
\uchyph=200
\brokenpenalty=200
\hbadness=10000
\clubpenalty=10000
\widowpenalty=10000
\displaywidowpenalty=10000
\pretolerance=10000
\tolerance=2000
\nobreak
\penalty 5000
\hyphenpenalty=5000
\exhyphenpenalty=5000
\sequentialequations
 
      \let\elrm=\elevenrm

\let\elrm=\elevenrm

  \def\str{\penalty-10000\hfilneg\ } 
 
\def\doeack{\footnote{\star}{\leftskip=12pt\hsize 14.4truecm\elrm
            Work supported by Department of Energy contracts
DE--AC02--76ER03069 (MIT),
DE--FG05--84ER40215 (OSU) and   DE--AC03--76SF00515 (SLAC).}}
 
 
\REF\kane{G.L. Kane \etal, \plb {\bf B354} (1995) 350.}
 
\REF\ext{
M. Bastero-Gil, B. Brahmachari, \prd {\bf D54} (1996) 1063.\nextline
B. Brahmachari, R. Mohapatra, Int. J. Mod. Phys. {\bf A11} (1996) 1699.
\nextline
J. Ellis, J.L. Lopez, D.V. Nanopoulos, \plb {\bf B371} (1996) 65.
\nextline
J. K. Kim \etal, \prd {\bf D53} (1996) 1712.\nextline
P.H. Frampton \etal, preprint HEP-PH/9511445 (1995).}
 
\REF\shif{See \eg: M. Shifman, Mod. Phys. Lett. {\bf A10} No. 7 (1995)
605.}
 
\REF\pnb{P.N. Burrows, SLAC-PUB-7293; to appear in Proc. Int.
Symposium on Radiative Corrections, Cracow, 1996.}
 
 
\REF\bm{P.N. Burrows, H. Masuda, \zpc {\bf C63} (1994) 235.}
 
\REF\bmmo{P.N. Burrows \etal, \plb {\bf B382} (1996) 157.}
 
\REF\ert{R.K. Ellis, D.A. Ross, A.E. Terrano, Phys. Rev. Lett. {\bf 45}
(1980) 1226; Nucl. Phys. {\bf B178} (1981) 421.\nextline
G. Kramer, B. Lampe, Z. Phys. {\bf C39} (1988) 101; Fortschr.
Phys. {\bf 37} (1989) 161.}
 
\REF\kn{Z.~Kunszt \etal, CERN 89--08 Vol I, (1989) p. 373.}
 
\REF\opalalp{OPAL Collab., P.D. Acton \etal, \zpc {\bf C55} (1992) 1.}
 
\REF\sldalp{SLD Collab., K. Abe \etal, \prd {\bf D51} (1995) 962.}
 
\REF\pms{P.M. Stevenson, \prd {\bf D23} (1981) 2916.}
 
\REF\fac{G. Grunberg, \prd {\bf D29} (1984) 2315.}
 
\REF\blm{S.J. Brodsky, G.P. Lepage, P.B. Mackenzie, \prd {\bf D28}
(1983) 228.}
 
\REF\sam{M.A. Samuel, G. Li, E. Steinfelds, \prd {\bf D48} (1993) 869.}
 
\REF\detail{M.A. Samuel, G. Li, E. Steinfelds, \prd {\bf E51} (1995)
3911.}
 
\REF\pade{See \eg G.A. Baker Jr., `Essentials of \Pade Approximants',
Academic Press, 1975.}
 
\REF\RRtauBj{M.A. Samuel, J. Ellis, M. Karliner,
\prl {\bf 73} (1994) 1207.}
 
\REF\Bj{J. Ellis \etal, \plb {\bf B366} (1996) 268.}
 
\REF\hiro{H. Masuda, Y. Ohnishi, SLAC-PUB-6560 (1994); subm. to \prd D.}
 
\REF\event{EVENT program supplied by P. Nason.}
 
\REF\had{See \eg, Yu.L. Dokshitzer, G. Marchesini, B.R. Webber,
CERN-TH/95-281 (1995).}
 
\REF\pnbnlc{P.N. Burrows, SLAC-PUB-7093 (1996); To appear in Proc.
Workshop on Physics and Experiments with Linear Colliders,
September 8-12 1995, Morioka-Appi, Iwate, Japan.}

 
 
 
 
 
 
\frontpagetrue
 
 
{\hfill SLAC--PUB--7222}
 
{\hfill MIT-LNS-96-212}
 
{\hfill September 1996}
 
\vskip 1truecm
 
\title{\bf APPLICATION OF PADE APPROXIMANTS TO
DETERMINATION OF \alpmzsq FROM
HADRONIC EVENT SHAPE OBSERVABLES IN e$^+$e$^-$
ANNIHILATION\doeack}
 
\vskip 1truecm
 
\centerline{P.N. Burrows}
\centerline{Massachusetts Institute of Technology}
\centerline{Cambridge, MA 02139, USA}
 
\vskip .5truecm
 
\centerline{T. Abraha, M. Samuel, E. Steinfelds}
\centerline{Department of Physics}
\centerline{Oklahoma State University}
\centerline{Stillwater, OK 74078, USA}
 
\vskip .5truecm
 
\centerline{H. Masuda}
\centerline{Stanford Linear Accelerator Center}
\centerline{Stanford University, Stanford, CA 94309, USA}
 
\vskip 1truecm
 
\centerline{\bf ABSTRACT}
\noindent
{\elrm
We have applied \Pade approximants to perturbative QCD
calculations of event shape observables in \ep \ra hadrons.
We used the exact \oalpsq prediction and the [0/1] \Pade
approximant to estimate the \oalpc term for 15 observables,
and in each case determined \alpmzsq from comparison with
hadronic \z0 decay data from the SLD experiment.
We found the scatter among the \alpmzsq values to be significantly
reduced compared with the standard \oalpsq determination, implying that
the \Pade method provides at least a partial approximation of
higher-order perturbative contributions to event shape observables.
}
 
\vskip .7truecm
 
\centerline{\it Submitted to Physics Letters B}
 
\vfill
\eject
 
\doublespace
\parskip = 0pt
 
One of the most important tasks in high energy physics is
the precise determination of the strong coupling \alp, conventionally
expressed at the scale of the mass of the \z0 boson,
$M_Z$ $\simeq$ 91.2 GeV. Not only does measurement of \alpmzsq in
different hard processes and at different hard scales $Q$ provide a
fundamental test of the theory of strong interactions, Quantum
Chromodynamics (QCD), but it also allows constraints on
extensions to the Standard Model of elementary particles. For example,
it has been claimed [\kane] that measurements of electroweak observables
can be better described by the Standard Model with \alpmzsq = 0.112 and
the addition of light superpartners, than by the Standard Model alone
with \alpmzsq = 0.123. The latter study, as well as studies of other
possible extensions to the Standard Model [\ext],
have been prompted by claims [\shif] that
recent measurements of \alpmzsq may be grouped into two classes:
those made at
`low-$Q^2$', which tend to cluster at values around 0.112, and
those made at
`high-$Q^2$', which tend to cluster at values around 0.123; for a review
of these measurements see Ref.~[\pnb].
 
Examination of the large set of \alpmzsq measurements reveals that in
fact all are consistent with a `world average'
central value of about 0.117 with an uncertainty of $\pm0.005$
[\pnb], and that their grouping into two supposedly discrepant classes
is arbitrary and not significant. Furthermore, nearly all measurements
are limited by theoretical systematic uncertainties that derive from
lack of knowledge of higher-order perturbative QCD contributions, or of
non-perturbative effects, or both. The effects on \alpmzsq determinations
of such uncalculated contributions have been estimated using
\adhoc procedures, sometimes in different ways by different
experimental collaborations working in similar areas; see \eg
[\bm,\bmmo].
It appears to us premature to speculate on beyond-Standard Model
explanations of \alpmzsq measurements that, within large, dominant and
arbitrarily-estimated theoretical uncertainties,
are consistent with one another. We suggest that a more
rational approach is to reduce the size of the limiting
theoretical uncertainties which may, or may not,
be concealing new physics.
 
Here we consider hadronic event shape observables in \epa, for which
perturbative QCD calculations
have been performed exactly only up to second order in \alp
[\ert,\kn] and have been used extensively by
collaborations at the PETRA, PEP, TRISTAN, SLC, and LEP colliders for
measurement of \alpmzsq~[\pnb]. A large scatter among the
\alpmzsq values derived from different observables has been found
[\opalalp,\sldalp], which can in principle be accounted for
as an effect of the \apriori unknown
higher order contributions in perturbation theory.
A consensus has arisen among experimentalists to estimate the size of
this effect
for each observable from the renormalisation-scale ($\mu$) dependence
of the \alpmzsq values derived from fits of the \oalpsq
calculation to the data,
see \eg [\sldalp], and to quote a corresponding renormalisation
scale uncertainty on \alpmzsq.
Within such uncertainties the
\alpmzsq values from the different observables are found to be
consistent, but this (arbitrary) procedure naturally leads to a large
uncertainty on the final \alpmzsq averaged over all observables;
for example, using 15 observables the SLD Collaboration determined
[\sldalp]:
$$
\alpmzsq \quad = \quad 0.1226 \pm0.0025\;{\rm (exp.)} \pm 0.0109\;
{\rm (theor.)}
\eqno\eq
$$
where the theoretical uncertainty is dominated by the contribution
of $\pm 0.0106$ from the renormalisation scale variation.
 
The best resolution of this situation would be to calculate the
observables to higher order in perturbation theory, a difficult and
unattractive task that has not yet been achieved. In the absence of
\oalpc QCD calculations it has been suggested [\pms,\fac,\blm]
that the \oalpsq calculation for each observable
can be `optimised' by choosing a specific value of the
renormalisation scale. It has recently been shown [\bmmo] that such
optimised scale
choices do not serve to reduce the scatter among the \alpmzsq values
determined from different observables, which implies that they do not
reduce the size of the contributions from the uncalculated higher orders.
 
Here we employ an approach for estimating the \oalpc
contributions to, as well as the sum of, the perturbative QCD series for
event shape observables in \epa.
The method makes use of \Pade Approximants (PA). The PA method
has been applied to estimate coefficients in perturbative quantum field
theory and statistical physics [\sam] and
is outlined in detail in Ref.~[\detail]. We give a brief review here.
 
The PA $[N/M]$ to the series:
$$
S\quad=\quad S_0\;+\;S_1 x \;+\;S_2 x^2 \;+\; \ldots \; +\;
S_{N+M} x^{N+M}
\eqno\eq
$$
is defined [\pade]:
$$
[N/M]\quad\equiv\quad {a_0 + a_1 x + a_2 x^2 + \ldots + a_N x^N \over
{1 + b_1 x + b_2 x^2 + \ldots + b_M x^M}},
\eqno\eq
$$
where $N$ and $M$ are integers such that $N\geq 0$ and $M>0$, and
$$
[N/M]\quad=\quad S + O(x^{N+M+1}).
\eqno\eq
$$
The coefficients $a_i$ ($0\leq i\leq N$) and $b_j$ ($1\leq j\leq M$)
are obtained by multiplying Eq.~4 by the denominator of
Eq.~3 and equating coefficients of like powers of $x$.
By consideration of the terms of O$(x^{N+M+1})$ one can obtain an
estimate of the coefficient $S_{N+M+1}$. Furthermore,
for an asymptotic series $[N/M]$ can be taken to be an estimate of the
sum of the series to all orders; we refer to this as the \Pade sum (PS).
 
Recently this method has been applied to estimate the \oalpc term in
the perturbative QCD prediction for the inclusive
quantities $R$ = $\sigma (e^+e^-\rightarrow
$ hadrons)/$\sigma(e^+e^-\rightarrow\mu^+\mu^-)$, $R_{\tau}$ =
$\Gamma(\tau$\ra$\nu$ hadrons)/
$\Gamma(\tau$\ra$e\overline{\nu}_e\nu_{\tau})$ and the
Bjorken sum rule in deep inelastic scattering [\RRtauBj]. In each case
the PA estimate is in good agreement with the exact calculation of
the \oalpc term, which is remarkable since the method does not
involve knowledge of strong interaction dynamics. Based upon this
success PA estimates have also been made of \oalpf terms for the
same quantities [\RRtauBj,\Bj].
 
In the case of \epa into hadronic final states the perturbative
QCD prediction for an infra-red- and collinear-safe observable
$X$ can be written:
$$
{1\over \sigma} {d\sigma\over dX}(X,\mu)
=
\overline{\alpha_s}(\mu)A(X)+
\overline{\alpha_s}^2(\mu)\;B(X,\mu)+
\overline{\alpha_s}^3(\mu)\;C(X,\mu)+
O(\overline{\alpha_s}^4(\mu))
\eqno\eq
$$
where $\alpbar\equiv\alpha_s/2\pi$. To date only the leading and
next-to-leading coefficients $A(X)$ and $B(X,\mu)$ have been
calculated [\ert,\kn].
Throughout this paper we set the renormalisation scale $\mu$ to
the so-called physical value $\mu=Q=\sqrt{s}$;
in this case the $\mu$-dependence
of the beyond-leading coefficients $B$, $C$, $\ldots$ vanishes.
 
We have employed the 15 hadronic event shape
observables used in the recent \alpmzsq
determination by the SLD Collaboration [\sldalp], namely
1$-$thrust ($\tau$), oblateness ($O$), the $C$-parameter ($C$),
normalised heavy jet mass ($\rho$),
total jet broadening ($B_T$), wide jet broadening ($B_W$),
the differential 2-jet rate defined using the E, E0, P, P0, D and G
algorithms ($D_2^E$, $D_2^{E0}$, $D_2^P$, $D_2^{P0}$, $D_2^D$ and
$D_2^G$ respectively), energy-energy correlations ($EEC$) and their
asymmetry ($AEEC$), and the Jet Cone Energy Fraction ($JCEF$) [\hiro].
Distributions of these 15 event shape observables were
measured [\sldalp] using a sample of approximately 50,000
hadronic \z0 decay events.
The data were corrected for detector
bias effects such as acceptance, resolution, and inefficiency, as well
as for the effects of initial-state radiation and
hadronisation, to arrive at `parton-level'
distributions, which can be compared directly with the QCD calculations.
 
For each observable $X$ we employed the EVENT program [\event] to
calculate the coefficients $A(X)$ and $B(X)$
in Eq.~5. These coefficients are listed in Table 1
for a representative value of each observable. In each case
it can be seen that the next-to-leading
coefficient $B$ is typically an order of magnitude larger than the
leading coefficient $A$. At first sight this appears to make the
perturbative approach invalid; it should be noted, however, that
the ratio of next-to-leading to leading terms in Eq.~5 contains
an additional factor of \alpbar, which is about 1/52 for
\alpmzsq = 0.12,
so that in all cases the next-to-leading term is smaller than
the leading term.
 
By explicitly separating an overall factor of \alpbar on the r.h.s.
of Eq.~5 and comparing with the form of Eq.~2
the PA [0/1] can be defined for the series.
For each bin of each observable, by consideration of the PA $[0/1]$
we derived an estimate of the
coefficient $C(X)$ of the next-to-next-to-leading term
in the series:
$$
C(X)\quad=\quad{B(X)^2 \over{A(X)}}.
\eqno\eq
$$
Examples of the predictions for $C(X)$ are given in Table 1.
It follows from Eq.~6 that $C(X)\;\geq\;0$ and that the ratios
$C(X)/B(X)$ and $B(X)/A(X)$ are equal.
For each bin of each observable we added the PA prediction for
the \oalpc term to the exact \oalpsq calculation to obtain an estimate
of the series to \oalpc. For each observable we then
fitted these calculations simultaneously to all bins in the
selected range of the data [\sldalp]
by minimising $\chi^2$ w.r.t. variation
of \alpmzsq. The resulting \alpmzsq and $\chi^2_{dof}$ values
are shown in Table 2.
 
It should be noted that the \oalpc estimate does not provide a good fit
to the $B_T$ data, as indicated by the large $\chi^2_{dof}$ value
(Table~2). It is perhaps significant that this observable
shows a large renormalisation scale uncertainty on \alpmzsq
determined at \oalpsq [\sldalp], and has a very
large ratio of next-to-leading to leading order coefficients (Table 1),
both implying that beyond-next-to-leading order contributions may be
large. Furthermore, the latter suggests that
the PA [0/1] estimate of the next-to-next-to-leading order
coefficient may be neither reliable, nor sufficient to describe the data
as is observed. In the following discussion
we therefore exclude $B_T$ and restrict the analysis to the remaining
14 observables; however, we indicate in footnotes to the text
any relevant changes in results if $B_T$ is included.
 
The \alpmzsq values are shown in Fig.~1, together with the
corresponding values from the \oalpsq analysis of the SLD data with
$\mu=\sqrt{s}$ [\bmmo]. For each observable
it can be seen that the \alpmzsq value derived using the \oalpc
estimate is lower than that derived using the \oalpsq calculation,
which is expected since $C(X)$ is positive.
Also shown for each observable is the total uncertainty
on \alpmzsq from the SLD \oalpsq study [\sldalp], which is
dominated by the renormalisation scale variation. In each case
the \oalpc \alpmzsq value lies near the lower bound given by the scale
uncertainty on the \oalpsq result.
To the extent that the PA \oalpc estimate is accurate, this
implies that the uncertainty assigned to the \oalpsq \alpmzsq value
from each observable due to variation of the renormalisation scale is a
reasonable estimate of the effect of the missing \oalpc contribution.
 
Furthermore, it can be seen from Fig.~1 that
the scatter among the \alpmzsq values is noticeably smaller in
the \oalpc case.
Taking an unweighted average\footnote{1}{Weighted averages
using experimental errors [\sldalp]
yield mean \alpmzsq values consistent with the respective
unweighted averages within the statistical error on a single \alpmzsq
value.} and r.m.s. deviation over
each set of 14 \alpmzsq values yields
\footnote{2}{Including $B_T$ and averaging over 15 observables yields
$\alpmzsq = 0.1265\;\pm\;0.0076$ (\oalpsq) and
$\alpmzsq = 0.1139\;\pm\;0.0045$ (\oalpc).}:
$$
\alpmzsq \quad=\quad 0.1255\;\pm\;0.0070\qu \qu (\oalpsq)
\eqno\eq
$$
$$
\alpmzsq \quad=\quad 0.1147\;\pm\;0.0035\qu \qu (\oalpc),
\eqno\eq
$$
implying that the inclusion of \oalpc terms causes a reduction
in the central value of \alpmzsq at $\mu$ = $\sqrt{s}$
by approximately 0.011, and that the residual scatter among the \oalpc
\alpmzsq values,
presumably due to missing \oalpf terms, is approximately $\pm0.0035$,
which is comparable with the combined experimental error
and hadronisation uncertainty
on a single observable measured at $Q$ = $M_Z$ [\sldalp].
Since the accuracy of the \Pade Approximant method can only be verified
\apost, exact calculation of the \oalpc terms in order
to confirm these results would be extremely desirable.
 
We also used the PS [0/1] as an estimate of the sum of the
asymptotic series and extracted \alpmzsq by comparison with the data in
a similar manner. The \alpmzsq values are shown in
Fig.~1. Typically, for each observable, the PS \alpmzsq value is close
to the \oalpc value.
Again the fit to $B_T$ is very poor, and we omit it from the
following main discussion. Taking an unweighted average and r.m.s.
deviation over the set of 14 \alpmzsq values
yields\footnote{3}{Including $B_T$ and averaging over 15 observables
yields $\alpmzsq = 0.1136\;\pm\;0.0068$.}:
$$
\alpmzsq \quad=\quad 0.1148\;\pm\;0.0052\qu \qu ({\rm PS}).
\eqno\eq
$$
Though the average value is close to that obtained using the PA \oalpc
estimate, the r.m.s. deviation is somewhat larger,
implying that the PS [0/1] provides
a poorer estimate of the sum of the series than the PA [0/1] estimate
to \oalpc.
 
Other interesting features are apparent from Fig.~1. Several
observables yield noticeably larger scale uncertainty at \oalpsq,
namely $\tau$, $B_T$, $O$, $C$, $D_2^{E}$ and EEC, which is an
indication of potentially large contributions at \oalpc that is
supported by the PA and PS [0/1] estimates.
If these are omitted from consideration the remaining set of
observables
yields average and r.m.s. \alpmzsq values of
$0.1212\pm0.0044$ (\oalpsq analysis),
$0.1131\pm0.0028$ (\oalpc analysis) and
$0.1155\pm0.0025$ (PS analysis).
Though the selection of this subset is arbitrary, it is perhaps
noteworthy that both \Pade-derived \alpmzsq values are close to the
corresponding values [\bmmo] obtained with \oalpsq calculations and
`PMS' [\pms] and `FAC' [\fac] optimised scales,
$0.1146\pm0.0025$ and $0.1148\pm0.0025$ respectively,
and that the scatter is similarly small
\footnote{4}{Significantly larger scatter is obtained with
the `BLM' scale choice: \alpmzsq = $0.1082\pm0.0091$.}.
 
In summary, we have applied \Pade approximants to the determination of
\alpmzsq from hadronic event shape observables in \epa. We applied the
PA [0/1] to the \oalpsq perturbative QCD series for 15 observables
to obtain estimates of the \oalpc coefficients, and then fitted the
extended series to SLD data to determine \alpmzsq.
With the renormalisation scale fixed to the c.m. energy
the scatter (neglecting $B_T$) among the \alpmzsq values was
reduced from $\pm0.0070$ (\oalpsq) to $\pm0.0035$ (\oalpc),
and the central value of \alpmzsq was lowered by 0.011.
If the scatter is interpreted
as arising from missing higher-order contributions, the
reduction in scatter implies that the \Pade method provides at least a
partial approximation of higher-order perturbative QCD contributions to
event shape observables. Furthermore,
if the \Pade-estimated \oalpc terms are accurate, this result implies
that residual \oalpf terms
contribute to \alpmzsq at the level of $\pm0.0035$, which is
comparable with
current experimental and hadronisation uncertainties.
 
These results based on \Pade approximants are tantalising,
but they can only be verified upon completion of a full perturbative
QCD calculation at \oalpc, which we strongly encourage.
One could then apply the [0/2] or [1/1] \Pade
approximants to the \oalpc series in order to estimate the size of
\oalpf contributions.
Since the accuracy of the \Pade method in predicting
unknown perturbation series coefficients is expected to improve with
increasing order of the known terms, it can be argued that the
estimated \oalpf coefficients would be expected to be more accurate
than the estimates of \oalpc coefficients presented
here\footnote{5}{See Refs.~[\RRtauBj,\Bj] for \oalpf estimates
for $R$, $R_{\tau}$ and the Bjorken sum rule.}. Combined with recent
theoretical progress in understanding hadronisation effects in terms of
`power corrections' [\had], it may
then be plausible to expect that \alpmzsq measurements at the 1\%-level
of precision could be achieved at future high-energy
\ep colliders [\pnbnlc].
 
We thank D. Muller and Y. Ohnishi for helpful contributions, and also
S.~Brodsky, L.~Dixon and T.~Rizzo for comments on the manuscript.
P.N.B. thanks Lawrence Morland for useful discussions concerning \Pade
approximants.
 
\refout

\vfill\eject
 
\phantom{}
 
\vskip 1truecm
 
\begintable
Observable         | value $X$ | $A(X)$ | $B(X)$ | $B(X)/A(X)$ |
$C(X)$ \cr
$\tau$             |  0.18     | 23.93  | 711.1  | 29.7 | 21130
 \nr
$\rho$             |  0.15     | 36.55  | 624.5  | 17.1 | 10670
 \nr
$B_T$              |  0.23     | 21.69  | 1042   | 48.0 | 50020
 \nr
$B_W$              |  0.14     | 82.16  | 1402   | 17.1 | 23940
 \nr
$O$                |  0.21     | 71.27  | $-$363.0 | $-$5.09 | 1848
\nr
$C$                |  0.46     | 18.89  | 518.9  | 27.8 | 14260
 \nr
$y_c$ ($D_2^E$)    |  0.17     | 38.84  | 1217   | 31.3 | 38140
 \nr
$y_c$ ($D_2^{E0}$) |  0.17     | 38.84  | 822.1  | 21.2 | 17400
 \nr
$y_c$ ($D_2^P$)    |  0.12     | 78.10  | 1029   | 13.2 | 13550
 \nr
$y_c$ ($D_2^{P0}$) |  0.17     | 38.84  | 601.6  | 15.5 | 9318
\nr
$y_c$ ($D_2^D$)    |  0.12     | 33.10  | 490.6  | 14.8 | 7271
\nr
$y_c$ ($D_2^G$)    |  0.22     | 37.93  | 412.7  | 10.9 | 4489
\nr
$\chi$ ($EEC$)     |  91.8$^\circ$ | 2.460  | 44.23  | 18.0 | 795
\nr
$\chi$ ($AEEC$)    |  41.4$^\circ$ | 2.682  | 22.74  | 8.48 | 193
\nr
$\chi$ ($JCEF$)    | 131.4$^\circ$ | 5.196  | 50.97  | 9.81 | 500
\endtable
 
\vskip .5truecm
 
\noindent
Table 1.~
Perturbative QCD calculation of  $1/\sigma d\sigma/dX$ for each
observable (first column) at a representative value $X$ (second column).
The third and fourth columns show the leading and
next-to-leading order coefficients, respectively.
The ratio of coefficients is shown in the fifth column, and
the PA [0/1] prediction for the next-to-next-to-leading order
coefficient $C(X)$ is listed in the sixth column.
 
\vfill\eject
 
\phantom{}
 
\vskip 1truecm
 
\begintable
Observable | \alpmzsq | $\chi^2_{dof}$
\cr
$\tau$     | $0.1172\pm$0.0009 | 0.6
\nr
$\rho$     | $0.1183\pm$0.0007 | 0.9
\nr
$B_T$      | $0.1031\pm$0.0009 | 112
\nr
$B_W$      | $0.1143\pm$0.0008 | 1.6
\nr
$O$        | $0.1230\pm$0.0011 | 0.6
\nr
$C$        | $0.1158\pm$0.0008 | 2.7
\nr
$D_2^E$    | $0.1164\pm$0.0006 | 6.4
\nr
$D_2^{E0}$ | $0.1105\pm$0.0007 | 3.5
\nr
$D_2^P$    | $0.1144\pm$0.0008 | 0.8
\nr
$D_2^{P0}$ | $0.1122\pm$0.0009 | 3.4
\nr
$D_2^D$    | $0.1156\pm$0.0011 | 4.0
\nr
$D_2^G$    | $0.1122\pm$0.0008 | 3.9
\nr
$EEC$      | $0.1154\pm$0.0008 | 0.4
\nr
$AEEC$     | $0.1082\pm$0.0012 | 0.4
\nr
$JCEF$     | $0.1124\pm$0.0007 | 0.2
\endtable
 
\vskip .5truecm
 
\noindent
Table 2.~Values of \alpmzsq and $\chi^2_{dof}$
from fits of QCD predictions
incorporating the PA [0/1] estimate for the \oalpc term.
The errors are statistical only and are highly correlated between
observables.
 
\vfill\eject
 
\noindent{\bf Figure Caption}
 
\vskip .5truecm
 
\noindent
FIG.~1. Values of \alpmzsq determined from fits to event shape
observables (see text): (a)
\oalpsq$\;$(circles); (b) \oalpc estimate (squares);
(c) \Pade sum (PS) (crosses). The shaded band shown
for each observable is dominated by
the renormalisation scale uncertainty on the \oalpsq fit.
For each point combined experimental statistical and systematic errors
[\sldalp] are shown; the errors are highly correlated between
observables.
 
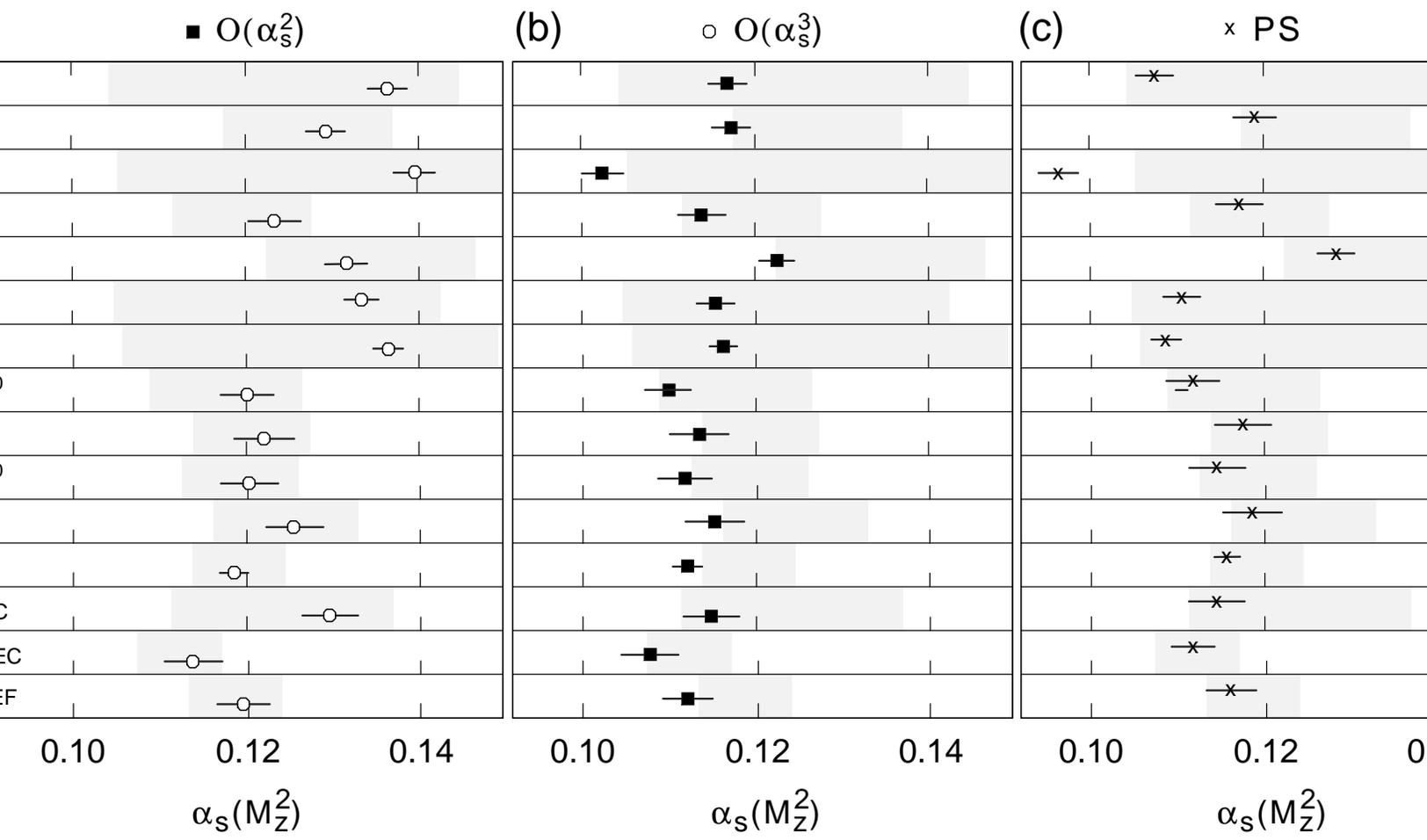
\bye